# Computational Studies of Light Shift in Raman-Ramsey Interference-Based Atomic Clock

G. S. Pati[1,*], Z. Warren[1], N. Yu[2] and M.S. Shahriar[3,4]

[1]Department of Physics and Engineering, Delaware State University, Dover, DE 19901
[2]Jet Propulsion Laboratory, California Institute of Technology, 4800 Oak Grove Drive, Pasadena, CA 91109
[3]Department of Electrical Engineering and Computer Science, Northwestern University, Evanston, IL 60308
[4]Department of Physics and Astronomy, Northwestern University, Evanston, IL 60308

[*]Corresponding author: gspati@desu.edu

Determining light shift in Raman-Ramsey interference is important for the development of atomic frequency standards based on a vapor cell. We have accurately calculated light shift in Raman-Ramsey interference using the density-matrix equations for a three-level system without invoking the adiabatic approximation. Specifically, phase shifts associated with coherent density-matrix terms are studied as they are relevant to the detection of Raman-Ramsey interference in transmission (or absorption) through the medium. For the single-velocity case, the numerically computed results are compared with the analytical results obtained using the adiabatic approximation. The result shows light shift suppression in conformity with the closed-form analytic solutions. The computational studies have also been extended to investigate Raman-Ramsey interference for a Doppler-broadened vapor medium. Importantly, a velocity-induced frequency shift is found at the fringe center as an additional source of frequency error for a vapor cell Raman clock.

## 1.    INTRODUCTION

In recent years, Raman-Ramsey (RR) interference has been extensively studied as a method for developing high performance atomic clocks [1-7]. Experimentally, RR interference is produced by creating Raman excitation in a three-level Λ-type atomic medium using two optical fields separated in either space or time. In this case, two-photon resonant Raman excitation by the optical fields creates atomic coherence via coherent population trapping (CPT) between the hyperfine ground states in the alkali atoms. Subsequently, RR interference is produced by virtue of an interaction-free evolution of the atoms between the separated optical fields. Since its inception, RR interference has been studied in different types of atomic media ranging from atomic beams and thermal vapor to cold atoms. Among these, vapor cell-based RR interference [8-10] is being studied with significant interest, particularly for developing compact atomic clocks with high frequency-stability performance.

When used as an atomic reference, RR interference offers the advantage of producing narrow-linewidth and high-contrast resonance without being sensitive to laser power-broadening. This accounts for improved short-term frequency stability of an RR clock which is directly proportional to the resonance linewidth, and inversely proportional to the signal contrast. The long-term frequency stability of the clock is usually dictated by the frequency shifts caused due to a number of factors such as laser frequency drift, laser intensity noise, variations in cell temperature, buffer gas pressure, external magnetic field, etc. [11,12]. A fundamental limit on the long-term stability is imposed by the light shift. Light shift is caused by electric-dipole interaction between the atoms and the optical fields. Typically, the largest variation in light shift is caused by the fluctuations and drifts in the laser intensity and laser detuning. The slope of the light shift as a function of laser intensity or laser detuning then affects the long-term stability of the clock.

The sensitivity of light shift to the laser parameters also depends on the type of excitation used in the clock. In many situations, it can be accurately calculated for predicting the performance of the clock. For example, in the case of an optically-pumped atomic clock, light shift-related frequency instabilities can be comprehensively modeled by investigating laser interaction with a two-level atomic system [11,13]. Similarly, light shift in a CPT clock can also be modeled using a three-level system interacting with the continuous bichromatic laser fields [12,14-17]. The complexity of this model can increase for a real atomic system which generally involves multiple-levels (e.g. hyperfine states and Zeeman sublevels) interacting with the bichromatic laser fields. Additionally, multiple sideband frequencies may need to be considered if a frequency-modulation technique is used to generate the beams in a CPT clock.

It is known that generation of RR interference relies on the CPT phenomenon. However, light shift properties of CPT are not directly applicable in the case of RR interference because of the pulsed excitation mechanism used in the RR interference. In a seminal paper, Hemmer *et al* [18] first obtained closed-form analytic expressions for RR light shift by applying an adiabatic approximation. Their results revealed an important, but possibly counter-intuitive property that the RR light shift can be reduced (or suppressed) by utilizing a strong interaction in the first zone. This prediction was verified by experiments using a sodium atomic beam [18]. Unlike an atomic beam experiment, RR interference in the atomic vapor can be influenced by several additional factors such as off-resonant excitations, rapid ground-state dephasing, and more significantly, by Doppler broadening. Presence of these factors can potentially change the properties of RR light shift for a vapor medium.

The aim of this study is to investigate RR light shift using a computational model based on the density-matrix equations for a generalized three-level system without invoking the adiabatic approximation originally used by Hemmer et al [18]. Eliminating the adiabatic approximation allows us to calculate RR light shift and its reduction to greater precision. Results using our model show the dependence of RR light shift on laser detuning and intensity. Results are also obtained to investigate RR phase shifts associated with coherent density-matrix elements as they are relevant to the detection of RR interference in transmission (or absorption). The paper is organized as follows: first, we describe the density-matrix equations to create a computational framework for studying RR interference. We then obtain analytical results for RR interference using the adiabatic approximation to compare them with our computational results. Finally, we extend our computational studies using the three-level model to investigate RR interference in a Doppler-broadened medium, and discuss the implications of velocity-averaging on the performance of a vapor cell RR clock.

## 2. THEORETICAL MODEL FOR RAMAN-RAMSEY INTERFERENCE

For our calculation, we consider a two-photon resonant Raman excitation in a closed three-level Λ-system as shown in Fig. 1a. Typically, in a rubidium-based Raman clock, Raman excitation is formed by using D1 ($\lambda = 795$ nm) optical transitions in $^{87}$Rb atoms. Fig. 1a shows a Raman excitation created by using the hyperfine levels of $5S_{1/2}$ and $5P_{1/2}$ states in $^{87}$Rb atoms. As the hyperfine states, $F' = 1$ and $F' = 2$ are separated by more than the Doppler linewidth in rubidium, a minimal three-level interaction model can be used to provide a good physical insight into RR interference and its light shift properties. A theoretical model describing the interaction of optical fields with the atoms, and the time-evolution of the system is formed by using the following Liouville density-matrix equations [19,20]:

$$\dot{\rho}_{11} = -\left(i\frac{\Omega_1}{2}\rho_{13} + c.c\right) + \Gamma_1 \rho_{33} - \gamma(\rho_{11} - \rho_{22})$$

$$\dot{\rho}_{22} = -\left(i\frac{\Omega_2}{2}\rho_{23} + c.c\right) + \Gamma_2 \rho_{33} + \gamma(\rho_{11} - \rho_{22})$$

$$\dot{\rho}_{33} = \left(i\frac{\Omega_1}{2}\rho_{13} + c.c\right) + \left(i\frac{\Omega_2}{2}\rho_{23} + c.c\right) - \Gamma\rho_{33}$$

$$\dot{\rho}_{13} = -i\frac{\Omega_1}{2}(\rho_{11} - \rho_{33}) - i\frac{\Omega_2}{2}\rho_{12} - \left(\Gamma_1 + \frac{\gamma}{2} + i\delta_1\right)\rho_{13}$$

$$\dot{\rho}_{23} = -i\frac{\Omega_2}{2}(\rho_{22} - \rho_{33}) - i\frac{\Omega_1}{2}\rho_{21} - \left(\Gamma_1 + \frac{\gamma}{2} + i\delta_2\right)\rho_{23}$$

$$\dot{\rho}_{12} = i\frac{\Omega_1}{2}\rho_{32} - i\frac{\Omega_2}{2}\rho_{13} - [\gamma + i(\delta_1 - \delta_2)]\rho_{12} \qquad (1)$$

where $\rho_{ii}$ and $\rho_{ij}$ ($i,j = 1,2,3$) are density-matrix elements that correspond to atomic population and coherence between the energy-levels, respectively, and $\sum_{i=1}^{3} \rho_{ii}$ is constant for a closed system. The rotating wave approximation and the rotating wave transformation have been used in the above equations. The equations also include $\gamma$ as the rate of collision induced incoherent population exchange between the ground states $|1\rangle$ and $|2\rangle$, and $\Gamma = (\Gamma_1 + \Gamma_2)$ as the total decay of atomic population from $|3\rangle$. The terms $\Omega_1 [= (\mu_{13}.\mathcal{E}_1)/\hbar]$ and $\Omega_2 [= (\mu_{23}.\mathcal{E}_2)/\hbar]$ are defined as the Rabi frequencies for the two Raman beams with $\mu_{12}$ and $\mu_{23}$ being defined as dipole-matrix elements for $1 \leftrightarrow 3$ and $2 \leftrightarrow 3$ transitions, respectively, and ($\mathcal{E}_1$ and $\mathcal{E}_2$) as electric-field amplitudes for the Raman beams. Frequency detunings $\delta_1$ and $\delta_2$ correspond to the two single-photon detunings associated with the system. These can also be redefined in terms of average (or common-mode) detuning $\delta = (\delta_1 + \delta_2)/2$ and difference-frequency (or Raman) detuning $\Delta = (\delta_1 - \delta_2)$ of the two optical fields.

A resonant Raman excitation ($\Delta = 0$) in a $\Lambda$-system prepares the atoms in the medium in a non-absorbing 'dark' state defined by $|-\rangle = (\Omega_2|1\rangle - \Omega_1|2\rangle)/\sqrt{\Omega_1^2 + \Omega_2^2}$. This serves as a basis for observing CPT resonance in a $\Lambda$-system. Theoretically, properties of CPT resonance can be studied by obtaining steady-state solutions [19] to the set of density-matrix eqn. (1). Unlike CPT, RR interference is observed using a pulsed interrogation technique. Fig. 1b shows a typical pulse sequence used in experiments and is also used here in our model to generate RR interference [4]. This involves Raman interaction with a long pulse, known as the CPT pulse of duration $\tau_c$, followed by a free-evolution time T, and another Raman interaction with a short pulse known as the query pulse of duration $\tau_q$, shown in Fig. 1b. In RR interference, the CPT pulse first prepares the atoms in the dark state. The interference effect is created by the query pulse due to an accumulated phase difference $\Delta\varphi\ (= \Delta \cdot T)$ between the atomic coherence and the evolving phase of the Raman fields during the time interval T.

A steady-state analysis cannot be used to investigate RR interference. Time-dependent density-matrix eqn. (1) needs to be solved for three successive time intervals $(\tau_c, T, \tau_q)$ in order to describe the evolution of all density-matrix elements and the dynamics of RR interference. It is, however, possible to obtain analytical solutions to eqn. (1) by applying an adiabatic approximation to the evolution of the excited state $|3\rangle$ [14,18]. This approximation is valid in cases where the decay rate $\Gamma$ is large compared to all other rates in the system (i.e. $\Gamma \gg \Omega_1, \Omega_2, \delta_1, \delta_2, \gamma$). Under this condition, the system can rapidly come in equilibrium with the ground states. In such a case, the system of eqn. (1) can be reduced to form an effective two-level system only consisting of the two ground-state matrix elements. Hemmer et al [18] has obtained an analytic solution for the excited state population $\rho_{33}$ which is given by

$$\rho_{33}(\tau_c + T + \tau_q) = \alpha\, e^{-\alpha\Gamma\tau_q}\left[1 - (1 - e^{-\alpha\Gamma\tau_c})\,|\sec(\varphi)|\cos(\Delta T - \varphi_{LS})\right] \quad (2)$$

where $\alpha = \Omega^2/(\Gamma^2 + 3\Omega^2 + 4\delta^2)$ is a parameter related to the CPT pulse saturation, and $\Omega^2 = (|\Omega_1|^2 + |\Omega_2|^2)/2$ is the square of the average Rabi frequency, and is proportional to the intensity of the CPT pulse. As seen in eqn. (2), RR interference is generated by changing the difference-detuning, $\Delta$. It is also found from eqn. (2) that the center of RR interference is shifted from $\Delta = 0$ by a phase angle, $\varphi_{LS}$, due to light shift (or AC Stark shift), which is given by

$$\Delta\nu_{LS} = \frac{\varphi_{LS}}{2\pi T} = \frac{1}{2\pi T}\tan^{-1}\left[\left(\frac{e^{-\alpha'\Gamma\tau_c}}{1 - e^{-\alpha\Gamma\tau_c}}\right)(\rho_{11} - \rho_{22})^0 \sin(\alpha'\delta\tau_c)\right] \quad (3)$$

where $(\rho_{11} - \rho_{22})^0$ corresponds to the initial population difference between the ground states before excitation, and $\alpha' = \Omega^2/(\Gamma^2 + 4\delta^2)$. This expression is obtained for equal Rabi frequencies (i.e. $\Omega_1 = \Omega_2 = \Omega$) of the Raman beams in the CPT pulse. It suggests that interaction with the CPT pulse is responsible for creating light shift in RR interference. The arctangent relation also implies that RR light shift can be reduced (or suppressed) by choosing a sufficiently strong interaction with the CPT pulse ensuring the condition $\alpha'\Gamma\tau_c \gg 1$ or, equivalently, by increasing the integrated CPT pulse intensity ($\Omega^2\tau_c$). Under ideal conditions, light shift can also be completely cancelled if $(\rho_{11} - \rho_{22})^0$ or $\delta$ is equal to zero.

Frequency shift produced due to RR light shift can be made small by simply increasing the free-evolution time T which, in practice, is limited by the ground-state decoherence $\gamma$ ($T < 1/\gamma$). The above discussed properties of RR light shift can be graphically illustrated by using a Bloch vector representation [19,21]. In a vapor medium, RR interference is usually detected in transmission (or absorption) as opposed to the fluorescence. Absorption is described by the coherent density-matrix terms $\text{Im}(\rho_{13})$ and $\text{Im}(\rho_{23})$, which are given by

$$\text{Im}\begin{pmatrix}\rho_{13}\\ \rho_{23}\end{pmatrix} = \pm\alpha' e^{-\alpha'\Gamma(\tau_c+\tau_q)}(\rho_{11} - \rho_{22})^0 \cos(\alpha'\delta\tau_c)\left[\frac{\left(\delta \pm \frac{\Delta}{2}\right)}{\Omega}\sin(\alpha'\delta\tau_q) - \frac{\Gamma}{2\Omega}\cos(\alpha'\delta\tau_q)\right]$$

$$-\frac{\Gamma}{2\Omega}\alpha'(1-3\alpha)e^{-\alpha\Gamma\tau_q} + \left(\frac{\Gamma}{2\Omega}\right)\alpha'\left(1-e^{-\alpha\Gamma\tau_c}\right)e^{-\alpha\Gamma\tau_q}(1-3\alpha)|\sec(\varphi)|\cos(\Delta T - \varphi_{LS})$$

$$\mp \alpha'(1-e^{-\alpha\Gamma\tau_c})e^{-\alpha'\Gamma\tau_q}\left[\frac{\left(\delta\pm\frac{\Delta}{2}\right)}{\Omega}\cos(\alpha'\delta\tau_q) + \frac{\Gamma}{2\Omega}\sin(\alpha'\delta\tau_q)\right]|\sec(\varphi)|\sin(\Delta T - \varphi_{LS})$$

(4)

Unlike RR interference described in eqn. (2), eqn. (4) is found to contain oscillating quadrature signals (third and fourth terms). One can simplify these two terms in eqn. (4) to find an effective phase shift at the fringe center individually for $Im(\rho_{13})$ and $Im(\rho_{23})$, which are given by

$$\begin{pmatrix}\varphi_{13}\\\varphi_{23}\end{pmatrix} = \pm \tan^{-1}\left[\frac{(2\delta \pm \Delta)\cos(\alpha'\delta\tau_q) + \Gamma\sin(\alpha'\delta\tau_q)}{\Gamma(1-3\alpha)}e^{-(\alpha'-\alpha)\Gamma\tau_q}\right] \quad (5)$$

The phase shifts $\varphi_{13}$ and $\varphi_{23}$ are created by non-zero laser detuning $\delta$ and finite duration of the query pulse. If the interference is detected using the absorption of only one of the two individual Raman beams (while both Raman beams are used in the CPT zone as well as the query zone), these phase shifts will manifest in RR interference as $\cos\left[\Delta T - \varphi_{LS} - \begin{pmatrix}\varphi_{13}\\\varphi_{23}\end{pmatrix}\right]$. On the other hand, if RR interference is detected via absorption of both Raman beams in the query zone, described by the term $Im(\rho_{13} + \rho_{23})$, eqn. (4) shows that one of the oscillating quadrature signals cancels due to the fact that $\varphi_{13}$ and $\varphi_{23}$ have opposite signs and nearly equal values for $\Delta \ll (\delta, \Gamma)$. Thus, RR interference detected using both Raman beams will only be affected by the light shift $\varphi_{LS}$ similar to the interference described by $\rho_{33}$. Eqn. (5) also provides information regarding the dependence of phase shifts $\varphi_{13}$ and $\varphi_{23}$ on laser intensities, detuning, and pulse parameters, etc. Next, we will discuss the dynamics of pulsed interrogation using our computational model.

## 3. RESULTS AND DISCUSSIONS

We solved density-matrix equations described in eqns. (1) for pulsed excitation conditions to investigate the properties of RR interference. A computational eigenvalue method has been used for solving the density-matrix equations. The following initial conditions were used prior to the CPT pulse: $\rho^0_{ij\,(i\neq j)} = \rho^0_{33} = 0$, $(\rho_{11} + \rho_{22})^0 = 1$, and $(\rho_{11} - \rho_{22})^0 = \rho^0$, where $\rho^0$ is assigned a non-zero value to create unequal initial population between the ground-states. Solutions describing $\rho_{33}(\tau_c + T + \tau_q)$, $Im[\rho_{13}(\tau_c + T + \tau_q)]$, and $Im[\rho_{23}(\tau_c + T + \tau_q)]$ are obtained by solving eqns. (1). RR interference is generated by repeating the computation for different values of $\Delta$ around $\Delta = 0$. Also, all frequency parameters involved in the system of eqns. (1) are normalized with respect to $\Gamma$. Phase shift in the RR fringe is determined by measuring the frequency shift at the fringe center, and using the frequency width of the fringes. For our studies, a phase shift measurement accuracy of about 1.57 mrad has been achieved by choosing the step-size in $\Delta$ to be 250 mHz, and fringe-width $\Delta\nu_{RR} = 500$ Hz. One can also compute an integrated solution (as in experiments [4,8]). For example, $\int_{\tau_c+T}^{\tau_c+T+\tau_q}\rho_{33}dt$ represents an integrated fluorescence over the duration, $\tau_q$, of the query pulse. Although this helps in increasing the SNR in the detection of RR fringes, the phase shift measured from an integrated solution is found to be the same as the one obtained using the instantaneous values. Hence, the computational results presented here are obtained using the instantaneous values of the density-matrix elements.

We will first discuss the phase shifts associated with RR interference by considering only single (or zero)-velocity group atoms, and compare them with the analytic results described in Section 2. Fig. 2 shows the phase shift measured using $\rho_{33}$ as a function of the average laser frequency detuning $\delta$. The result is obtained by using different average Rabi frequencies (or intensities) of the CPT pulse, and by considering equal intensities for the Raman beams in the CPT pulse. The following pulse parameters were used: $\tau_c = 200$ µs, T = 1 ms and $\tau_q = 100$ ns. An initial ground-state population difference corresponding to $\rho^0 = 0.2$ has been used. The result illustrates reduction in phase/light shift due to pulse saturation (i.e. $\Omega \tau_c \gg 1$) for higher light intensities. The result also shows the comparison between the computational results (solid line) and the analytic results obtained using adiabatic approximation (dashed line). The computational results estimate RR light shift more accurately than those obtained using the adiabatic approximation. RR light shift gets adequately suppressed for $\Omega \geq \Gamma/20$ over a range of $\delta$ close to $-\Gamma \leq \delta \leq \Gamma$. Light shift in the unsaturated region is found to oscillate (between positive and negative values), exhibiting nonlinear dependence on $\delta$. These oscillations can be physically explained by the spiraling of the Bloch vector around its steady-state value in a Bloch vector model [19,21]. The results in Fig. 2 also provides information regarding the long-term frequency stability of a RR clock limited by the variation in $\delta$, which can be potentially caused by sources such as laser drift, jitter, and linewidth broadening.

We have also studied the effect of decoherence due to the incoherent collisional exchange of population between the two ground-states on the RR light shift. The effect of this decoherence is found to be only pronounced in the region where light shift was previously suppressed under a zero-decoherence condition, and more so, for unequal Rabi frequencies (i.e. $\Omega_1 \neq \Omega_2$) of the CPT pulse. Fig. 3 shows RR light shifts calculated by choosing $\Omega_1 = \Gamma/5$ and $\Omega_2 = 0.75\,\Omega_1$ ($\Omega = \Gamma/5.65$), and by changing the exchange rate $\gamma$ from zero to $10^{-4}\Gamma$. Although the light shift still remains quite well suppressed over the range $-\Gamma \leq \delta \leq \Gamma$, the slope of light shift is found to increase by nearly three times with increasing $\gamma$. Choosing higher values of $\gamma$ produces significantly modified light shift and, simultaneously, a reduction in RR fringe contrast or signal-to-noise ratio (SNR). The fringe-width $\Delta \nu_{RR}$ observed in RR interference is also ultimately limited by $\gamma$ which, in experiments, is made small (i.e. $\gamma \ll \Gamma$) by using buffer-gas loaded cells.

Fig. 4 shows the computed RR light shift as a function of $\Omega$ of the CPT pulse. RR light shift is close to zero over the whole range of $\Omega$ when $\delta$ is set equal to zero. This also agrees with the prediction from eqn. (3). For large, non-zero values of $\delta$ (e.g. $\delta = \pm\Gamma$), oscillation in light shift with $\Omega$ is observed presumably due to weak excitation with the CPT pulse. At higher $\Omega$ (or intensity) of the CPT pulse, amplitudes of these oscillations quickly diminish by virtue of the strong interaction. Similar oscillations (shown by dashed lines) are also created by the presence of the term $\sin(\alpha'\delta\tau_c)$ in eqn. (3). The result once again shows that RR light shift can be significantly small under strong excitation conditions. Considering mono-velocity atoms and assuming a saturation intensity, $I_{sat}$, of 3 mW/cm$^2$, one can use this result to calculate RR light shift as a function of average intensity, $I \simeq (\Omega/\Gamma)^2 I_{sat}$, of the CPT pulse. One can also estimate the slope of RR light shift in the unsaturated or saturated regions by using this plot. Fig. 5 is a color-coded map generated using our computation to illustrate the properties of RR light shift as a function of both the average detuning, δ, and intensity, I, of the CPT pulse. The uniformly colored region in the plot where the pulse saturation condition ($\Omega\tau_c \gg 1$) is satisfied, corresponds to a small (close to zero) value of RR light shift. The plot also shows oscillations in light shift (as color change) at lower intensities below 500 μW/cm$^2$ and for $4 \leq |\delta| \leq 18$ MHz. The range of δ and I chosen in the map can arise in the experiment due to possible laser drift, linewidth broadening, lock instability and/or intensity noise, etc. Therefore, it can be used to get a first-hand estimate of the light shift and long-term frequency stability of the RR clock. For example, the slope of RR light shift can be reduced to an extremely small value below 0.001 Hz mW$^{-1}$ cm$^2$ by using a strong excitation condition. Assuming that laser detuning can be tightly controlled by the electronic servo, a long-term stability better than 10$^{-13}$ can be achieved in the RR clock by limiting the total intensity change in the laser to less than 1%.

The results presented so far describe RR light shift by using the excited-state population term $\rho_{33}$. This term only gives a measure of fluorescence intensity for describing RR interference produced in an atomic beam experiment. We have also calculated RR light shift using the coherence terms $\text{Im}(\rho_{13}), \text{Im}(\rho_{23})$, and $\text{Im}(\rho_{13} + \rho_{23})$ which are associated with RR interference detected using absorption/transmission of the beams. Fig. 6 shows the phase shifts associated with the coherent terms, plotted as a function of δ. In this particular case, the phase shift, $\varphi_{LS}$, due to RR light shift is suppressed to a negligibly small value because of pulse saturation since Ω is set equal to Γ/5. However, large phase shifts are still produced in $\text{Im}(\rho_{13})$ and $\text{Im}(\rho_{23})$. This is in sharp contrast to the phase shift observed in the population term $\rho_{33}$. The origin of these phase shifts were predicted earlier by the analytic theory represented by eqn. (5). The dashed line shows the analytic result in comparison with the computed results. In both cases, a large slope with respect to δ is observed. These phase shifts increase with δ, change sign around δ = 0 and become large (close to ± π/2) near $\delta = \pm \Gamma$. It is important to note that since the phase shifts measured in $\text{Im}(\rho_{13})$ and $\text{Im}(\rho_{23})$ are exactly opposite in sign, they cancel each other in the term $\text{Im}(\rho_{13} + \rho_{23})$ as shown in Fig. 6. These results establish the fact that phase shift observed in RR interference is dependent on how the fringe is detected using the transmitted Raman beams. The frequency error caused by this kind of phase shift in a cell-based RR clock can be cancelled by detecting RR fringes via the combined absorption of both Raman beams in the query zone.

We have generalized our computational model to investigate the RR phase/light shift in case of Doppler-broadening due to atomic motion. This is done by modifying the single-photon detuning terms in eqn. (1) as $\delta'_{1,2} = \delta_{1,2} \pm \omega_{1,2}(v/c)$ for the individual velocity-groups, and performing a weighted-average of the RR fringes calculated for each individual velocity-group. The weighting is a 1D Maxwell-Boltzmann velocity distribution described by $\left(1/\sqrt{2\pi}\, v_p\right) \exp\left(-v^2/v_p^2\right)$, where $v_p = \sqrt{2k_B T/m}$ is the most probable velocity corresponding to atomic mass, m, and sample temperature, T. First, we show the effect of velocity-averaging on the CPT resonance which is created using our model by choosing an interaction with a long CPT pulse of duration $\tau_c$ and setting T and $\tau_q$ to zero. As shown in Fig. 7, velocity-averaging reduces the linewidth of CPT resonance. This rather counter-intuitive result can be understood as follows. First, note that because the Raman beams are co-propagating, the differential Doppler shift is very small. Second, the linewidth for a CPT resonance is determined predominantly by the rate of optical pumping into the dark state. This rate is largest for the zero-velocity atoms, given by $\Omega^2\Gamma/$

$2(\Gamma^2 + 2\Omega^2)$ and is reduced for non-zero velocity atoms by a factor of $(\Omega^2 + \Gamma^2)/(\Gamma^2 + \Omega^2 + 4\delta_{DS}^2)$ where $\delta_{DS}$ is the common-mode Doppler shift for these atoms.

A Doppler-broadening close to a frequency-width 1 GHz is considered to simulate the effect of thermally-distributed atoms in a warm rubidium vapor cell. The line-shape of CPT resonance is calculated using the $\rho_{33}$ term. The calculation showed that, unlike CPT, line-shape of RR fringes remain virtually unchanged after velocity-averaging, because of the fact that the differential Doppler broadening is negligibly small. Note that the time separation, T, is a fixed parameter, and is the same for all atoms. This is in contrast to the space-separated RR interference produced in an atomic beam, where velocity averaging can cause rapid decay of fringe amplitude as T varies significantly with atomic velocity.

Fig. 8 shows computationally measured RR fringe shift obtained in the case of velocity-averaging. For comparison, the figure also shows the phase shifts measured for two non-zero, single-velocity groups corresponding to $k \cdot v = \pm 0.2\Gamma$ and $\pm \Gamma$, respectively. Velocity-averaging is found to produce a phase shift (slope $\simeq 0.12$ μrad.Hz$^{-1}$ for Rb) that increases linearly with δ. Even though this is an undesirable effect, the result shows a behavior which is expected from a Doppler-broadened medium. The following physical argument supports this result: individual velocity-group atoms experience different velocity-induced phase shifts since the Raman detuning, Δ gets modified as $\Delta' \to \Delta - (k_1 - k_2)v$ and $k_i = \omega_i/c$, where i = 1,2 due to the differential Doppler shift caused by the frequency mismatch between the Raman beams. If $\delta = 0$, this creates symmetrically displaced RR fringes produced by non-zero velocity-groups around the RR fringe produced by the zero-velocity group. Therefore, the resultant sum (or average) of these individual velocity-group responses does not create any net phase shift for $\delta = 0$. However, if $\delta \neq 0$, the fringe displacements become asymmetric around the zero-velocity group and, therefore, velocity-averaging creates a net velocity-induced phase shift. Such a phase shift linearly increases with δ and also changes sign with δ. The magnitude of this phase shift is also found to be large compared to the original light shift.

The velocity-induced phase/frequency shift found here is not specific to RR interference. It will manifest in any resonance including CPT that involves a resonant Raman excitation in a Doppler-broadened vapor medium. Velocity-induced shift does not depend on the laser intensity, and can cause a systematic frequency error in a Raman clock provided the uncertainty in laser detuning, δ can be minimized by a precise laser lock. We are currently working towards experimentally validating the presence of velocity-induced shift in a vapor cell-based RR clock. Additionally, we are also developing a multi-level computational model by including all magnetic sublevels for $^{87}$Rb atoms to further investigate how the properties of RR light shift established here get modified due to the presence of additional off-resonant excitations and collision-induced dephasing in the ground-state magnetic sublevels.

## 4. CONCLUSION

We have investigated light shift associated with Raman-Ramsey interference using a computational model based on a three-level atomic system. Light shift is accurately calculated without invoking the adiabatic approximation. We have also investigated the coherent density-matrix elements and the phase shifts associated with them, which is relevant for the case where RR interference is detection via absorption. The calculated results for the mono-velocity case confirm that RR light shift can be suppressed by orders of magnitude using the pulse saturation effect. Our study has also been extended to investigate light shift in the presence of Doppler-broadening in a medium. Although the light shift suppression remained valid in a Doppler-broadened medium by ensuring pulse saturation, additional velocity-induced phase/frequency shift is found to be associated with RR fringe center. Based on our computational results, we also predict that this will create a newly recognized source of frequency error for all vapor cell-based Raman clocks. As our results show, high long-term stability (below $10^{-13}$) in cell-based RR clock can be achieved by suppressing the light shift, and simultaneously reducing the impact of velocity-induced shift by a precise laser lock.

## ACKNOWLEDGMENTS

The authors acknowledge the support received from the DoD grant #W911NF-13-10152, the NASA URC grant #NNX09AU90A, and AFOSR grant # FA9550-10-1-0228 for conducting this research.

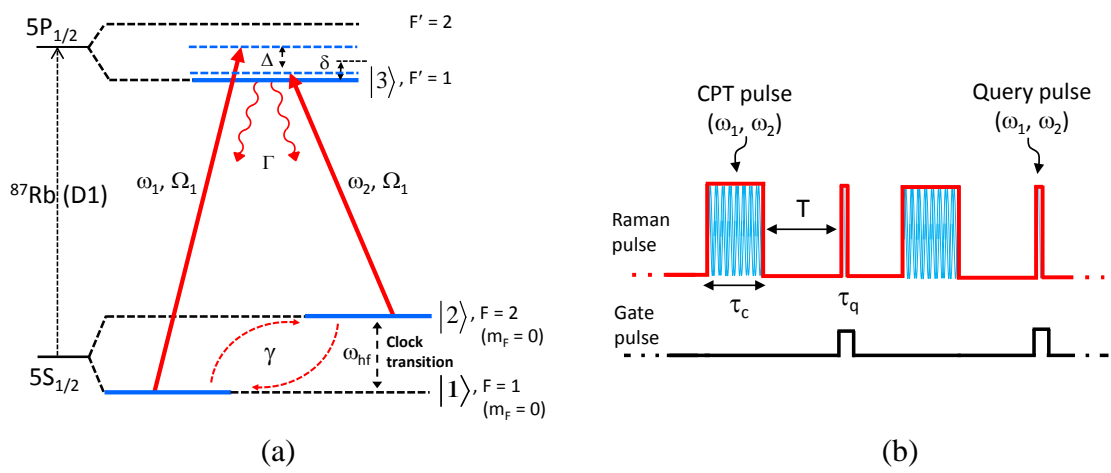

Fig. 1. (a) Raman excitation in a three-level Λ-system in $^{87}$Rb atoms, and (b) Raman pulse sequence considered in the computational model for generating RR interference.

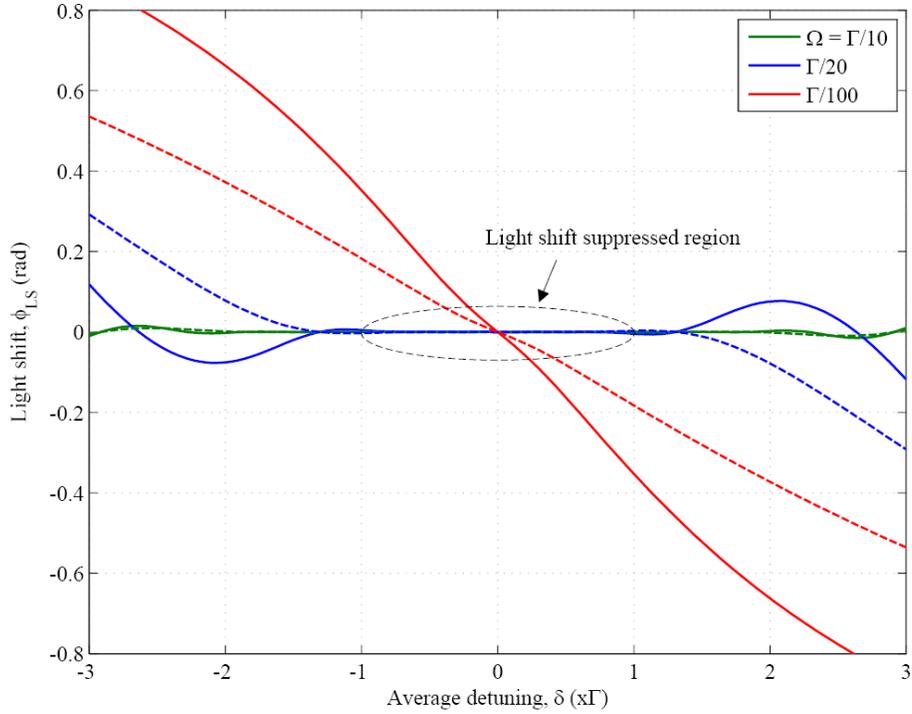

Fig. 2. Light shift, $\varphi_{LS}$ versus $\delta$ calculated using $\rho_{33}$. Solid lines represent numerically computed values, and dashed lines show analytical values given in eqn. (3). The following parameters were used in our calculations: $\tau_c = 200$ μs, $T = 1$ ms, $\tau_q = 100$ ns, $\Omega_1 = \Omega_2$, $(\rho_{11} - \rho_{22})^0 = 0.2$ and $\gamma = 0$. Encircled region shows the region of light shift suppression.

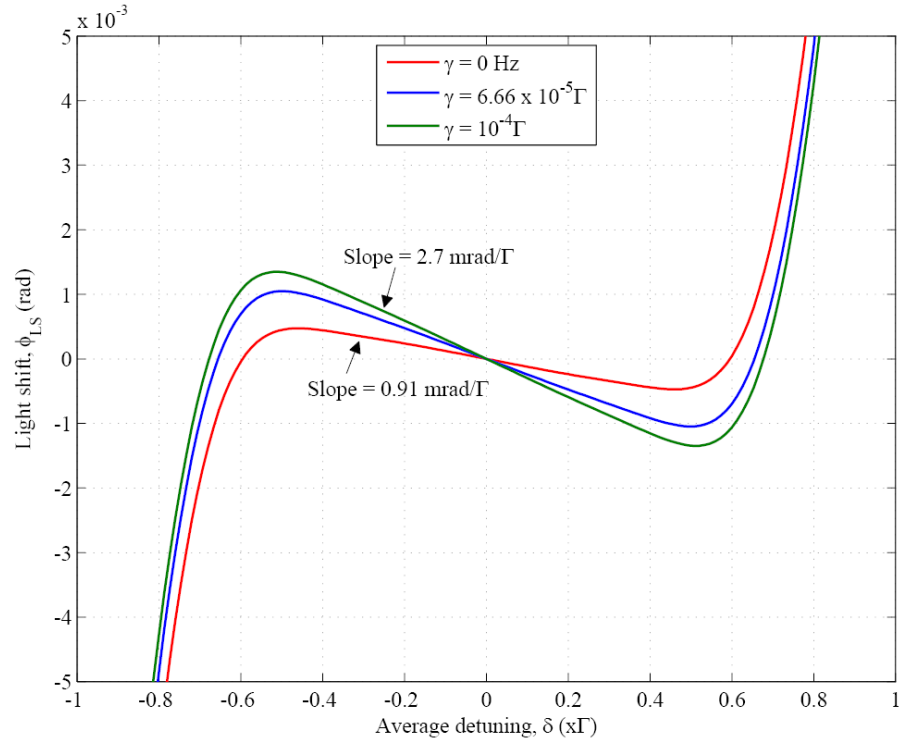

Fig. 3. Effect of decoherence $\gamma$ on the light shift, $\varphi_{LS}$. The following parameters were used in our calculations: $\tau_c = 200$ µs, $T = 1$ ms, $\tau_q = 100$ ns, $\Omega_1 = \Gamma/5$, $\Omega_2 = 0.75\,\Omega_1$, and $(\rho_{11} - \rho_{22})^0 = 0.2$.

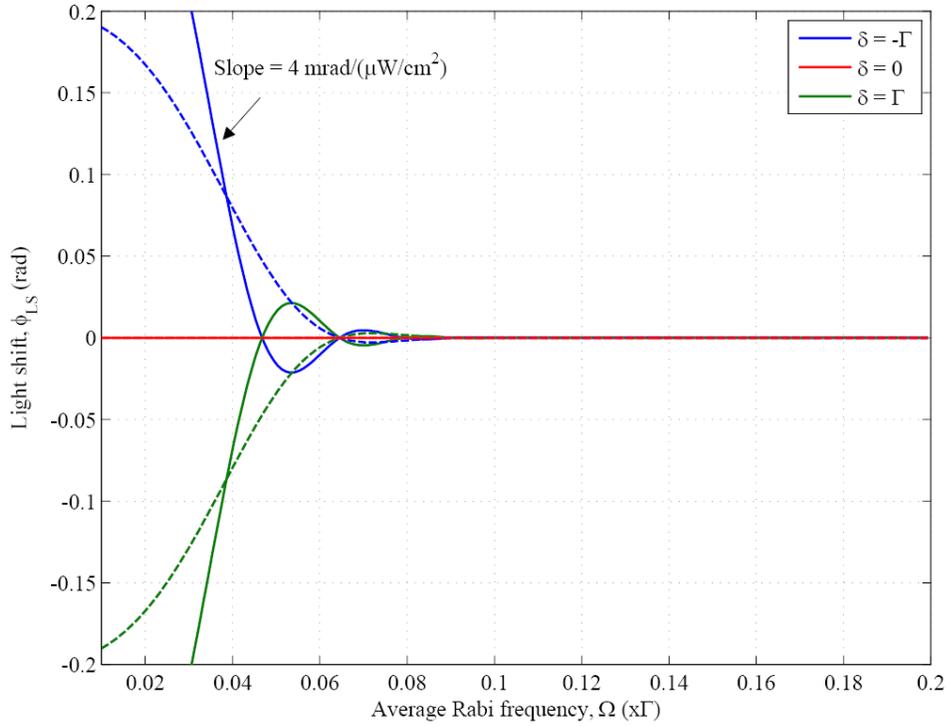

Fig. 4. RR light shift calculated as a function of the average Rabi frequency, $\Omega$, of the CPT pulse. Solid lines show numerical values, and dashed lines show analytical values obtained using eqn. (3). The following parameters were used in our calculations: $\tau_c = 100$ μs, $T = 1$ ms, $\tau_q = 100$ ns, $\Omega_1 = \Omega_2 = \Omega$, $(\rho_{11} - \rho_{22})^0 = 0.2$ and $\gamma = 0$. For $\delta = 0$, both numerical and analytical results show zero light shift over the entire range of intensity as expected.

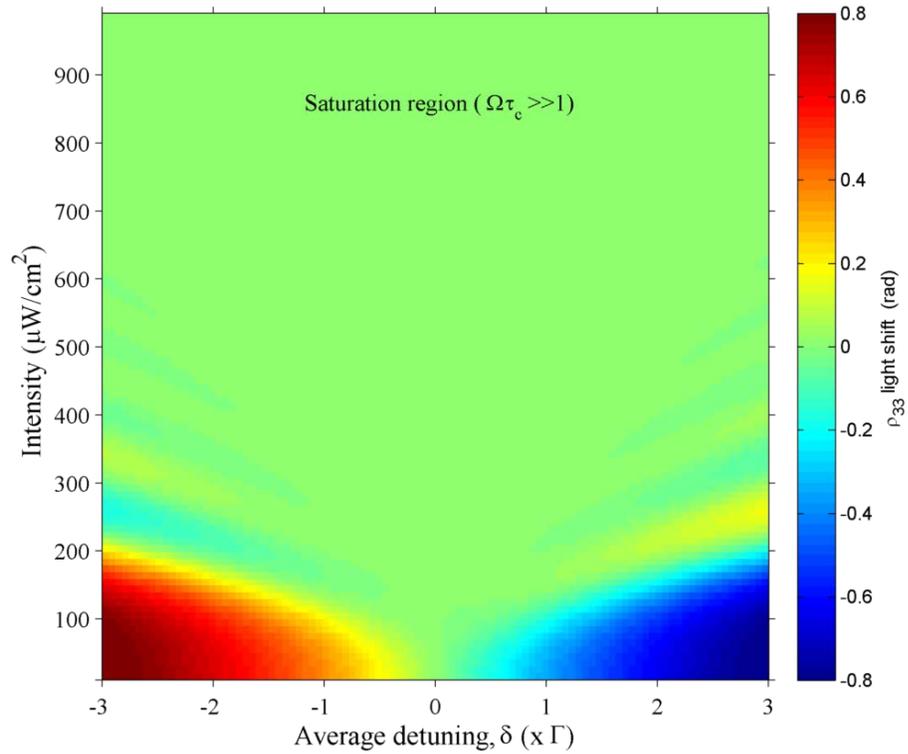

Fig. 5. Color-coded 2D plot showing light shift, $\varphi_{LS}$ as functions of average detuning, $\delta$ and intensity, I of the CPT pulse. Saturation intensity ($I_{sat}$ = 3 mW/cm$^2$) for a rubidium atom is considered in converting $\Omega$ to an intensity axis. The following parameters were used in our calculations: $\tau_c$ = 100 µs, T = 1 ms, $\tau_q$ = 100 ns, $\Omega_1 = \Omega_2$, $(\rho_{11} - \rho_{22})^0 = 0.2$ and $\gamma = 0$.

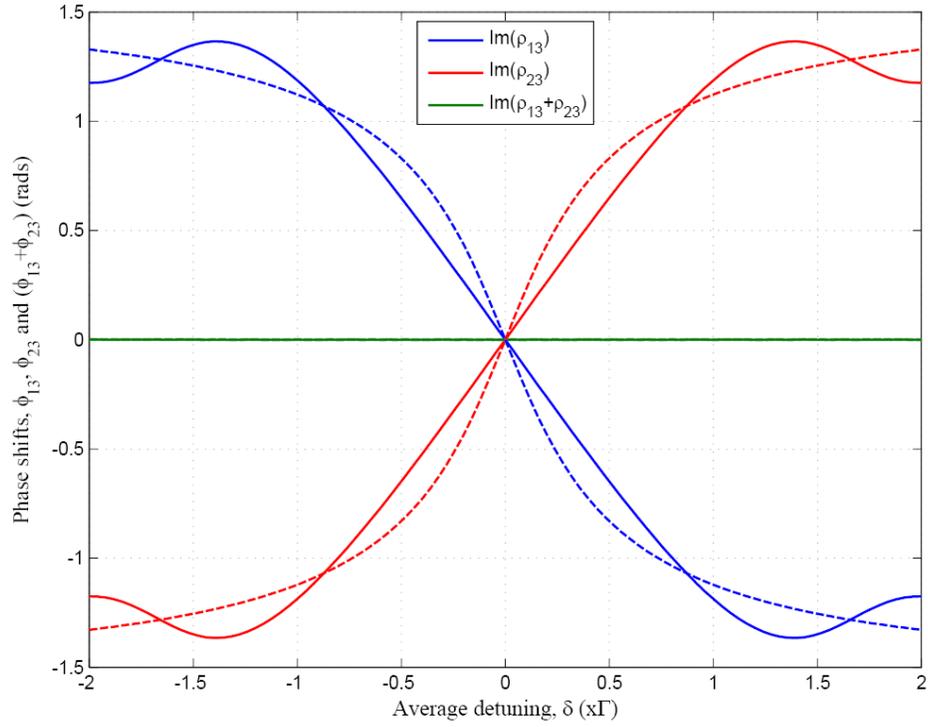

Fig. 6. Phase shifts associated with $\text{Im}(\rho_{13})$, $\text{Im}(\rho_{23})$, and $\text{Im}(\rho_{13} + \rho_{23})$ plotted as a function of $\delta$. Solid lines represent numerical values, and dashed lines show analytical results. The measured phase shifts cancel each other in $\text{Im}(\rho_{13} + \rho_{23})$. The following parameters were chosen in this calculation: $\tau_c = 200$ μs, $T = 1$ ms, $\tau_q = 100$ ns, $\Omega_1 = \Omega_2 = \Gamma/5$, $(\rho_{11} - \rho_{22})^0 = 0.2$ and $\gamma = 0$.

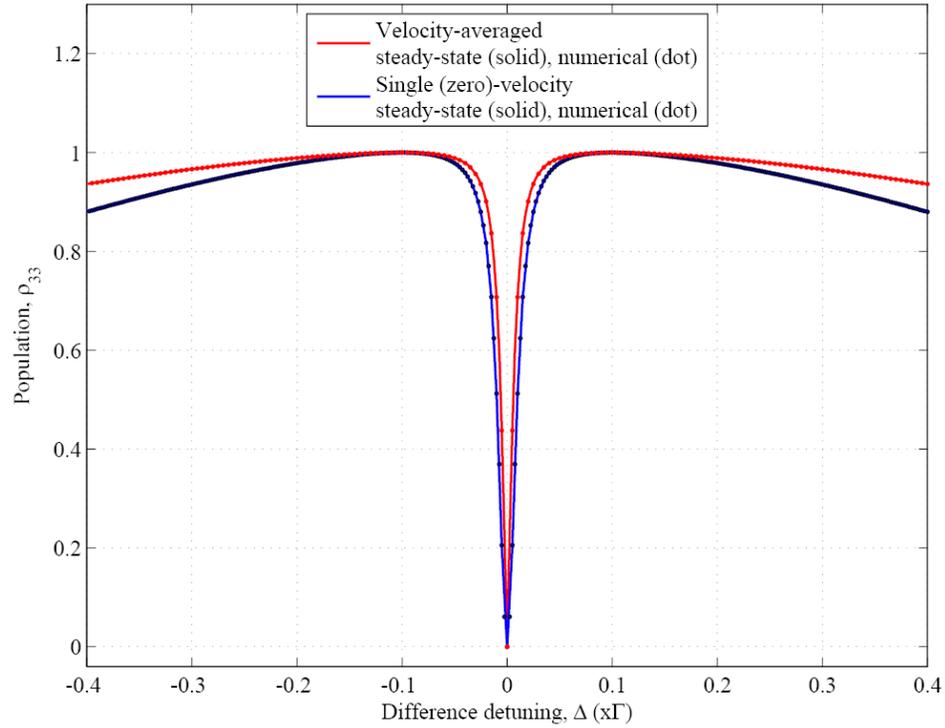

Fig. 7. Line-shape of CPT resonance plotted as a function of $\Delta$. Solid lines represent calculated line-shapes using the solutions to steady-state density-matrix equations with and without velocity-averaging, respectively. Filled dots show the match with our computational model using a long CPT pulse, $\tau_c = 2000$ µs, and parameters, $T = \tau_q = 0$, $\delta = 0$, $\Omega_1 = \Omega_2 = \Gamma/5$ and $\gamma = 0$. The resonance dip is normalized to show effective narrowing due to velocity-averaging.

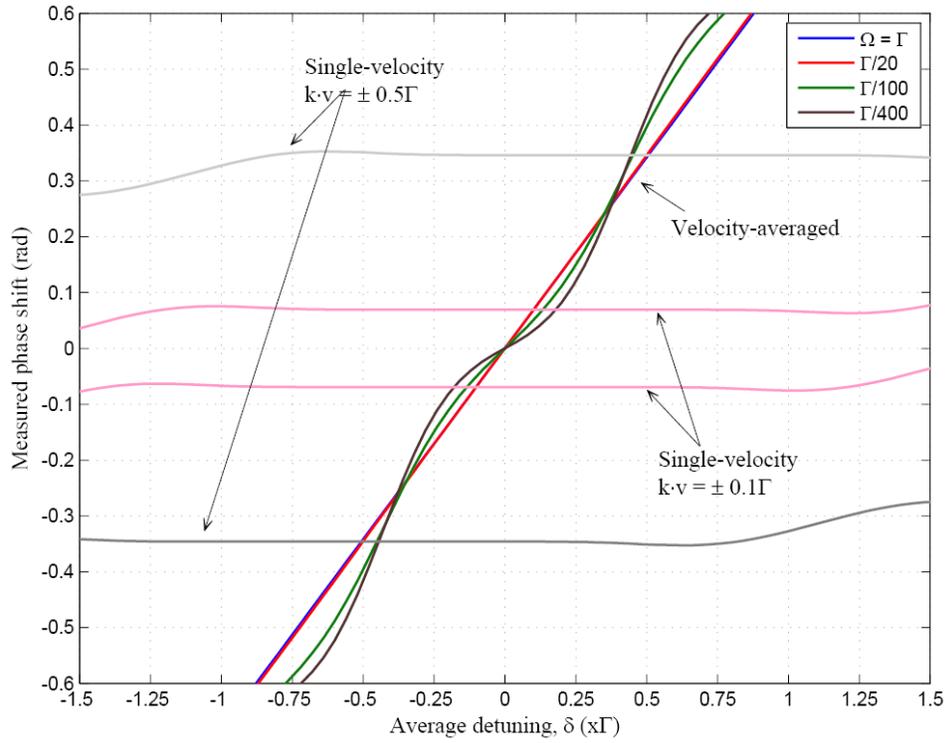

Fig. 8. Phase shift calculated in a Doppler-broadened medium and plotted as a function of δ. The following parameters were used in this calculation: $\tau_c = 200$ μs, $T = 1$ ms, $\tau_q = 100$ ns, $\Omega_1 = \Omega_2$, $(\rho_{11} - \rho_{22})^0 = 0.2$, and $\Delta\nu_{Doppler} \simeq 1$ GHz. Phase shifts for two single-velocity groups ($kv = \pm 0.1\Gamma$ and $\pm 0.5\Gamma$) are also shown in the figure corresponding to the case $\Omega_1 = \Omega_2 = \Gamma/20$.